\documentclass[11pt]{amsart}

\usepackage[colorlinks = true,
            linkcolor = red,
            urlcolor  = magenta,
            citecolor = black,
            anchorcolor = blue]{hyperref}
\usepackage{color}
\usepackage[shortalphabetic,initials,nobysame]{amsrefs}
\usepackage{upgreek}
\usepackage{tikz}
\usepackage[applemac]{inputenc} 
\usetikzlibrary{arrows}
\usepackage{xcolor}
\usepackage[all]{xy}
\usepackage{graphicx}
\usepackage{wrapfig}
\usepackage{yfonts}
\usepackage{graphicx}
\usepackage{amssymb}
\usepackage{epstopdf}
\usepackage{mathrsfs}
\usepackage[mathcal]{eucal}
\usepackage{bm}

\renewcommand{\eprint}[1]{\href{https://arxiv.org/abs/#1}{#1}}

\BibSpec{article}{%
    +{}  {\PrintAuthors}                {author}
    +{,} { \textit}                     {title}
     +{,} {}                            {note}
    +{.} { }                            {part}
    +{:} { \textit}                     {subtitle}
    +{,} { \PrintContributions}         {contribution}
    +{.} { \PrintPartials}              {partial}
    +{,} { }                            {journal}
    +{}  { \textbf}                     {volume}
    +{}  { \PrintDate}                {date}
    +{,} { \issuetext}                  {number}
    +{,} { \eprintpages}                {pages}
    +{,} { }                            {status}
    +{,} { \DOI}                   {doi}
    +{,} { \eprint}        {eprint}
      +{,} {\publisher}                {publisher}
    +{,} { \address}                   {address}
    +{}  { \parenthesize}               {language}
    +{}  { \PrintTranslation}           {translation}
    +{;} { \PrintReprint}               {reprint}
    +{.} {}                             {transition}
    +{}  {\SentenceSpace \PrintReviews} {review}
}

\newtheorem{theorem}{Theorem}[section]

\newtheorem{definition}[theorem]{Definition}

\newtheorem{Thm}{Theorem}[section]

\newtheorem{Con}[Thm]{Conjecture}

\theoremstyle{definition}
\newtheorem{Def}[Thm]{Definition}
\theoremstyle{remark}

\newtheoremstyle{named}{}{}{\itshape}{}{\bfseries}{.}{.5em}{#1 #3}
\theoremstyle{named}

\def\g{\mathfrak{g}}
\def\Frenkel:2013uda{\mathfrak{h}}

\def\bo{\textbf{o}}

\def\=>{\Longrightarrow}

\def\to{\longrightarrow}

\def\o+{\oplus}
\def\bo+{\bigoplus}

\def\<{\langle}
\def\>{\rangle}
\def\({\left(}
\def\){\right)}

\def\^{\wedge}
\def\+{\dagger}

\def\dd[#1,#2]{\frac{d#1}{d#2}}
\def\del[#1,#2]{\frac{\partial #1}{\partial #2}}
\def\over[#1]{\overline{#1}}
\def\vec[#1]{\overrightarrow{#1}}

\makeatletter
\def\mr@ignsp#1 {\ifx\:#1\@empty\else #1\expandafter\mr@ignsp\fi}%
\newcommand{\multiref}[1]{\begingroup
\xdef\mr@no@sparg{\expandafter\mr@ignsp#1 \: }%
\def\mr@comma{}%
\@for\mr@refs:=\mr@no@sparg\do{\mr@comma\def\mr@comma{,}\ref{\mr@refs}}%
\endgroup}
\makeatother

\newcommand{\hypref}[2]{\ifx\href\asklFrenkel:2013udaas #2\else\href{#1}{#2}\fi}
\newcommand{\Secref}[1]{Section~\multiref{#1}}
\newcommand{\secref}[1]{Sec.~\multiref{#1}}

\tikzset{->-/.style={decoration={
  markings,
  mark=at position .5 with {\arrow{latex}}},postaction={decorate}}}
\tikzset{
    >=latex
    }

\newcommand{\ard}{{\mathbf{d}}}

\newcommand{\nc}{\newcommand}
\nc{\on}{\operatorname}
\nc{\la}{\lambda}
\nc{\wh}{\widehat}
\nc{\ghat}{\wh\g}
\nc{\mb}{\mathbf}

\begin{document}
\title{Quantum K-theory and Integrability}

\author[P. Koroteev]{Peter Koroteev}
\address{
Department of Mathematics,
University of California,
Berkeley, CA 94720, USA
and 
\newline
Beijing Institute for Mathematical Sciences and Applications,
Beijing Huairou District, 101408, China
}
\date{\today}

\numberwithin{equation}{section}

\begin{abstract}
In this note, we explore some recent advancements in enumerative algebraic geometry, focusing particularly on the role of quantum K-theory of quiver varieties as viewed through the lens of integrable systems. We highlight a number of conjectures and open questions.
{\it This is a contribution to the proceedings of the GLSM@30 conference, which was held in May 2023 at the Simons Center for Geometry and Physics.}
\end{abstract}

\maketitle


\section{Introduction}
The connection between enumerative algebraic geometry and integrable systems has been recognized for decades. Beginning with the pioneering works of Witten and Dubrovin, this interplay was further developed in the 1990s through the contributions of Givental and his collaborators. More recently, advances in the understanding of supersymmetric gauge theories have merged with developments in geometric representation theory. In particular, the study of {\it symplectic resolutions} from a representation-theoretic perspective has revitalized this relationship, as demonstrated in the influential works of Okounkov and collaborators \cites{Braverman:2010ei,Okounkov:2015aa,2012arXiv1211.1287M}. A notable result is the emergence of integrable systems, such as the XXX and XXZ models based on quantum groups, from enumerative geometry for a broad class of algebraic varieties, including {\it Nakajima quiver varieties} \cites{Nakajima:1999hilb,Ginzburg:}.

Modern enumerative algebraic geometry explores the properties of moduli spaces, focusing on the computation of intersection numbers, such as counting curves within projective varieties. In particular, equivariant quantum K-theory investigates Euler characteristics of properly defined coherent sheaves $\chi(X)$ as representations of the automorphism group associated with the problem. Many enumerative problems acquire a representation-theoretic interpretation when the equivariant pushforward of the virtual structure sheaf is identified with elements of an algebra acting on $X$ by correspondences \cite{CFK}. Examples of such algebras include the loop group $U_\hbar(\widehat{\mathfrak{gl}}_n)$, the double affine Hecke algebra (DAHA), and the Elliptic Hall algebra.

In enumerative geometry, standard objects include appropriate deformations of the cup product and tensor product in equivariant cohomology and K-theory, respectively. These deformations are parameterized by {\it K\"ahler parameters}, with coefficients determined by curve counting. 

The physics results of Nekrasov and Shatashvili \cite{Nekrasov:2009uh} have led to a conjecture about equivariant quantum K-theory: the quantum multiplication by the generating function for the exterior algebra of the tautological bundle coincides with the Baxter $Q$-operator for the Heisenberg XXZ spin chain. Moreover, since tautological bundles generate the entire quantum K-theory, the equivariant quantum K-theory ring can be described as the ring of symmetric functions of Bethe roots.

Cohomological enumerative counts have shown profound connections with integrable systems, such as the Toda lattice \cite{2001math8105G} and the elliptic Calogero-Moser model \cite{Negut_2009}. These discoveries inspired the study of K-theoretic enumerative counts and their connections to finite-difference integrable models like the $q$-Toda and Ruijsenaars-Schneider models \cites{Koroteev:2018azn,Koroteev:2021tl}. By leveraging algebraic and geometric methods \cites{Gaiotto_2013,Koroteev:2021tl}, new dualities have been uncovered between these integrable systems and quantum spin chains (XXX, XXZ, XYZ family), often referred to as \textit{quantum/classical dualities}. More recently, \cites{KSZ,Frenkel:2020} established a direct connection between many-body systems and spin chains through the study of \textit{(q-)opers} on $\mathbb{P}^1$.

\vskip.1in

This note is organized as follows. In \Secref{Sec:ClassicalInt} we review the necessary details of classical many-body systems, primarily, the trigonometric RS model. Then in \Secref{Sec:QuantumIntEnum} we introduce one-dimensional quantum integrable spin chains with periodic boundary conditions by means of quantum groups. This will be followed by the discussion of the quantum Knizhnik-Zamolodchikov equation whose solutions will lead us to the quasimap counts to Nakajima quiver varieties. Studying different stability conditions for quiver varieties will naturally lead us to the space of q-opers in \Secref{Sec:Opers}. Finally, \Secref{Sec:OpenProblems} addresses a few follow up questions from the main discussion, in particular the limit when the character of the fiber dilation $\hbar$ is sent to infinity \secref{Sec:CompatLimit}, quantum integrability of open spin chains and related algebro-geometric constructions in \secref{Sec:OpenChainsSpSO}, and new exciting applications to number theory in \secref{Sec:NumberTheory}.

\subsection*{Acknowledgements} I am grateful to my long-term collaborators A. Zeitlin, E. Frenkel, and D. Sage as well as to other colleagues for the stimulating discussions over the past years. I would also like to thank the organizers of GLSM\@30 Workshop at Simons Center for Geometry and Physics and Physics at Stony Brook in May 2023 where a presentation on the matters of this note was delivered.

\section{Classical Integrability}\label{Sec:ClassicalInt}
Calogero in 1971 introduced a new integrable system on $N$ particles which interact with each other pairwise via a rational potential. Moser in 1975 proved its integrability using Lax pair formalism. The resulting Calogero-Moser (CM) system has three different versions depending on the periodicity of its coordinates -- rational, trigonometric, and elliptic. CM systems also admit relativistic generalizations which are known as Ruijsenaars-Schneider (RS) systems and which can also be rational, trigonometric, and elliptic (see Fig. 1 in \cite{Koroteev:2023ab}).

The trigonometric RS system, or tRS, will be the main focus of this section. It can be introduced algebraically as follows \cite{Oblomkov2004}
\begin{Def}
Let $V$ be an $N$-dimensional vector space over $\mathbb{C}$. Let $\mathcal{M}'$ be the subset of $GL(V)\times GL(V)\times V\times V^\ast$ consisting of elements $(M,T, u,v)$ such that
\begin{equation}\label{eq:FlatConNew}
\hbar M T -  T M = u\otimes v^T\,.
\end{equation}
The group $GL(N;\mathbb{C})=GL(V)$ acts on $\mathcal{M}'$ by conjugation
\begin{equation}\label{eq:SimilarityTRansf}
(X,Y, u,v)\mapsto (gMg^{-1},gTg^{-1}, gu,vg^{-1})\,,\qquad g\in GL(V)\,.
\end{equation}
The quotient of $\mathcal{M}'$ by the action of $GL(V)$ is called Calogero-Moser space $\mathcal{M}$.
\end{Def}

Note that when $\hbar$ is not a root of unity the $GL(V)$ action above is free so $\mathcal{M}'$ and $\mathcal{M}$ are nonsingular.

In the basis where $M$ is a diagonal matrix with eigenvalues $\chi_1,\dots, \chi_N$ the components of matrix $T$ are given by the following expression:
\begin{equation} \label{eq:lax1}
T_{ij} = \frac{u_i v_j}{\hbar \chi_i - \chi_j}\,.
\end{equation}

One can define the {\it tRS momenta} $p_i,\,i=1,\dots,N$  as follows:
\begin{equation}\label{eq:lax2}
p_i = - u_i v_i \frac{\prod\limits_{k \neq i} (\chi_i - \chi_k)}{\prod\limits_{k}\left(\chi_i -\chi_k \hbar\right)}\,.
\end{equation}

Using the above formula we can represent the components of matrix $T$ \eqref{eq:lax1} by properly scaling vectors $u$ and $v$:
\begin{equation}\label{eq:LaxFullFormula}
T_{ji}(\{p_i\}, \{\chi_i\})= \frac{\chi_j(1 - \hbar)}{\chi_j - \chi_i \hbar}\prod\limits_{k \neq j} \frac{\chi_j-\chi_k \hbar}{\chi_j - \chi_k}\, p_j 
=\frac{\prod\limits_{k \neq i}(\chi_j-\chi_k \hbar)}{\prod\limits_{k \neq j}(\chi_j - \chi_k)}\, p_j \,.
\end{equation}
Matrix $T$ above is known as the Lax matrix of the tRS model \cite{MR1329481}.

The coefficients of the characteristic polynomial of the Lax matrix are the tRS Hamiltonians
\begin{equation}
 \text{det}\left(u\cdot 1 -  T(\chi_i, p_i,\hbar) \right) = \sum_{k=0}^L (-1)^lH_k(\chi_i, p_i,\hbar) u^{n-k}\,,
\label{eq:tRSLaxDecomp}
\end{equation}

The corresponding energy relations or integrals of motion can be obtained by equating the above characteristic polynomials to $\prod_i (u-\xi_i)$ 
\begin{equation}
\sum_{\substack{\mathcal{I}\subset\{1,\dots,L\} \\ |\mathcal{I}|=k}}\prod_{\substack{i\in\mathcal{I} \\ j\notin\mathcal{I}}}\frac{\hbar\,\chi_i - \chi_j }{\chi_i-\chi_j}\prod\limits_{m\in\mathcal{I}}p_m = e_k (\xi_i)\,,
\end{equation}
where $e_l$ is the $l$-th elementary symmetric function.

\section{From Quantum Integrability to Enumerative Counts}\label{Sec:QuantumIntEnum}

\subsection{Quantum Knizhnik-Zamolodchikov Equation}
The intertwining operators for the quantum affine algebra $U_{\hbar}(\hat{\mathfrak{g}})$, also known as {\it conformal blocks}, satisfy certain difference equations known as quantum Kniznik-Zamolodchikov (qKZ) equations \cite{Frenkel:92qkz}. 

Explicitly, qKZ equations can be written as follows: 
difference equations 
\begin{eqnarray}\label{qkz}
\Psi(a_{i_1}, \dots, q a_{i_k}, \dots , a_{i_n}, \textbf{z})=H^{(q)}_{i_k}\Psi(a_{i_1}, \dots,, a_{i_n}, \textbf{z}),
\end{eqnarray}
where the solutions $\Psi$ take values in $\mathcal{H}$ and operators $H^q_i$ are expressed 
\begin{equation}
H^q_i = Z\cdot R_{i_k1}(a_{i_k}/a_1)\cdots R_{i_k k-1}(a_{i_k}/a_{k-1}) \cdot R_{i_k k-1}(q a_{i_k}/a_{k+1})\cdots R_{i_k i_n}(q a_{i_k}/a_{i_n})  
\end{equation}
in terms of products of the R-matrices
\begin{equation}
R_{ij}(a_i/a_j):V_i(a_i)\otimes V_j(a_j)\to V_j(a_j)\otimes V_i(a_i),
\end{equation}
satisfying the Yang-Baxter equation
\begin{equation}
R_{i j}(a_i/a_j) R_{i k}(a_i/a_k) R_{j k}(a_j/a_k) = R_{j k}(a_j/a_k) R_{i k}(a_i/a_k) R_{i j}(a_i/a_j)
\end{equation}
Here we have defined $\mathfrak{g}$ to be a simple Lie algebra and $\hat{\mathfrak{g}}_{k=0}=\mathfrak{g}[t^{\pm 1}]$,  be the corresponding loop algebra. The finite-dimensional modules $\{V_i\}$ of $\mathfrak{g}$ give rise to the evaluation modules $\{V_i(a_i)\}$, where $a_i$ stands for the value of the loop parameter $t$. These modules form a tensor category. In order to describe the integrable model, we choose a specific object in this category $$\mathcal{H}=V_{i_1}(a_{i_1})\otimes \dots \otimes V_{i_n}(a_{i_n}),$$ which is referred to as physical space and its vectors are called states. 
The equation can be schematically illustrated in figure 
\begin{figure}
\includegraphics[scale=.3]{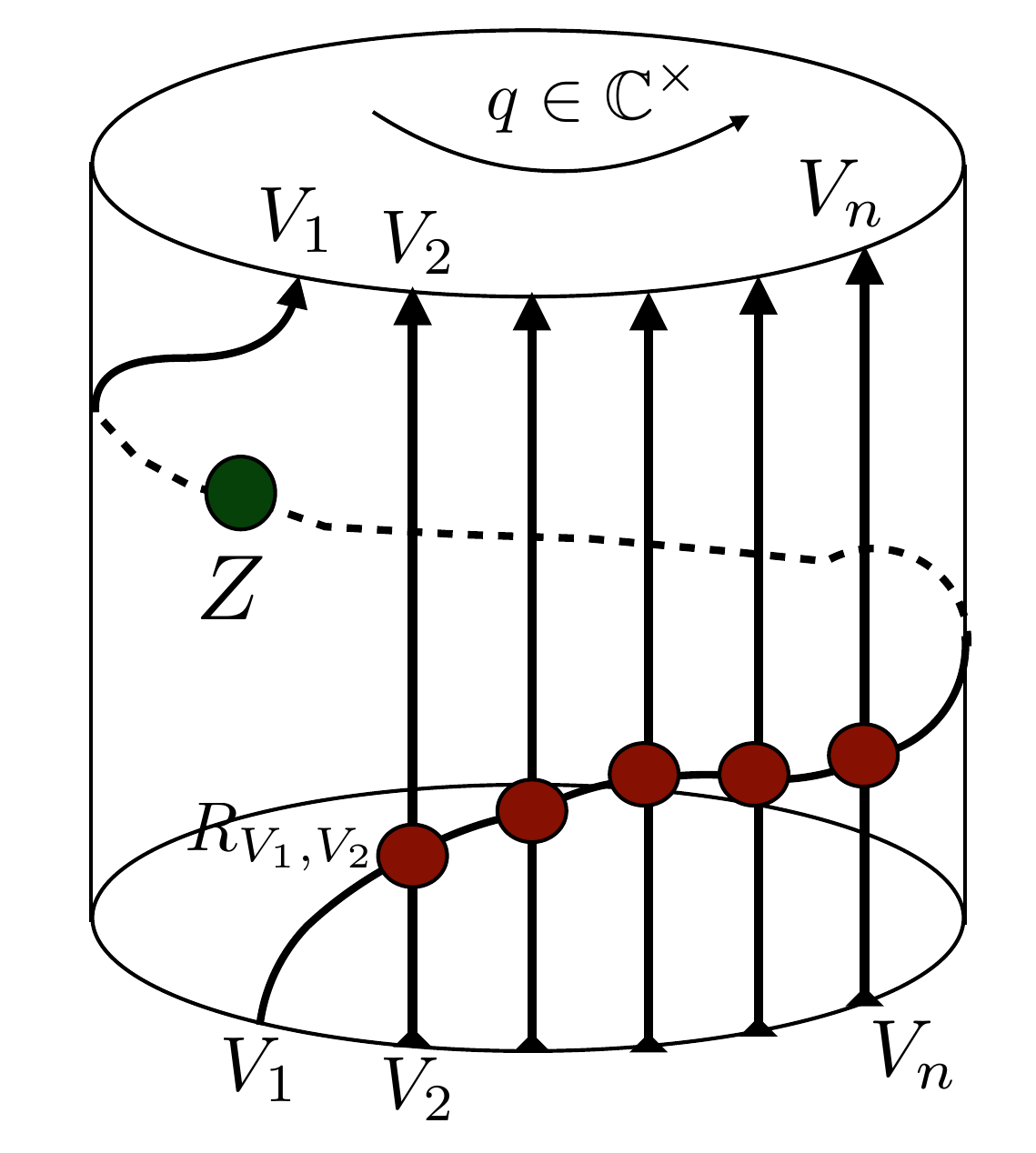}
\caption{Quantum KZ equation with twist element $Z$}
\end{figure}
For a given module $W(u)$ called the auxiliary module with spectral parameter $u$ we define the transfer matrix
\begin{equation}
\label{eq:TransfermatDef}
T_{W(u)}={\rm Tr}_{W(u)}\Big[ (Z\otimes 1)P{R}_{W(u),\mathcal{H}}\Big].
\end{equation}
Here the twist element is given by $Z=\prod^r_{i=1}z_i^{\check{\alpha}_i}\in e^{\mathfrak{h}}$, where $\mathfrak{h}$ is the Cartan subalgebra in $\mathfrak{g}$, $\{\check{\alpha}_i\}_{i=1,\dots, r}$ are the simple coroots of $\mathfrak{g}$, and  $P$ is a permutation operator. 
The Yang-Baxter equation implies that transfer matrices, corresponding to various choices of $W(u)$ form a commutative algebra, known as Bethe algebra. The commutativity of Bethe algebra implies integrability and the expansion coefficients of the transfer matrix yield nonlocal Hamiltonians of the XXX or XXZ spin chain depending on whether we deal with the Yangian or the quantum affine algebra respectively. 

The classic example of the closed Heisenberg XXZ spin chain corresponds to the quantum algebra $U_{\hbar}(\widehat{\mathfrak{sl}}(2))$ in which the physical space $\mathcal{H}$ is constructed from $V_i(a_i)=\mathbb{C}^2(a_i)$ -- the standard two-dimensional evaluation modules of $U_{\hbar}(\hat{\mathfrak{sl}}(2))$, while the twisted boundary conditions are described by $Z$.

\subsection{Solutions of qKZ}
The solution to the qKZ equation is given by an integral expression which schematically can be presented as \cite{Okounkov:2015aa,Aganagic:2017be}
\begin{equation}
\Psi_\alpha = \int_{\mathcal{C}} \frac{d\textbf{x}}{\textbf{x}} f_\alpha(\textbf{x},\textbf{a})\mathcal{K}(\textbf{x},\textbf{z},\textbf{a},q)
\end{equation}
with appropriately chosen contour $\mathcal{C}$. Here $\alpha$ is indexed by the physical space, while loop parameters $a$ are indexed by the representation. 
The universal kernel $\mathcal{K}$ has the following asymptotic behavior in the limit $q\rightarrow 1$: 
\begin{eqnarray} 
\mathcal{K} \simeq e^{\frac{Y(\textbf{x}, \textbf{z}, \textbf{a})}{\log(q)}}(\phi_0(\textbf{x}, \textbf{z}, \textbf{a})+\dots),
 \end{eqnarray}
In the limit $q\rightarrow 1$ the stationary phase approximation gives  $\Psi_\alpha=e^{\frac{S}{\log q}}(\Psi_{\alpha,0}+\dots)$, where $S=Y|_{\sigma_i},$  where $\sigma_i$ are the solutions of the equations $\partial_{x_i}Y=0$, which need to be solved with respect to $\textbf{x}$.  These equations coincide with the Bethe equations, and $\Psi_0$ is the eigenvector for the nonlocal Hamiltonians $H^{(1)}_i$. These operators emerge as coefficients from the expansion of the transfer matrices with respect to the spectral parameter, viz. $H^{(1)}_i\Psi_{\alpha,0}=p_i\Psi_{\alpha,0},~{\rm where} ~p_i=e^{\partial_{a_i}S}$. The latter coincides with the tRS momentum in the electric frame \eqref{eq:lax2}. 
Thus the map 
$$
\alpha \mapsto f_\alpha(\textbf{x}^*,\textbf{a})
$$
from the physical space to onshell Bethe eigenvectors provides diagonalization of the qKZ Hamiltonians. So one needs to find offshell functions $f_\alpha(\textbf{x},\textbf{a})$ first, which brings us to the enumerative algebraic geometry problem inspired by physics.

\subsection{Quiver Varieties}
Let us start with a framed quiver with set of vertices denoted by $I$. A representation of such quiver is given by a set of vector spaces $V_i$ corresponding to internal vertices and $W_i$ for the framing vertices, together with a set of morphisms between these vertices. Let $\text{Rep}(\mathbf{v},\mathbf{w})$ be the linear space of quiver representation with dimension vectors $\mathbf{v}$ and $\mathbf{w}$, where $\mathbf{v}_i=\text{dim}\ V_i$, $\mathbf{w}_i=\text{dim}\ W_i$, and let $\mu :T^*\text{Rep}(\mathbf{v},\mathbf{w})\to \text{Lie}{(G)}^{*}$, $G=\prod_i GL(V_i)$, be the moment map.

The Nakajima variety $X$ corresponding to a quiver is an algebraic symplectic reduction
\begin{equation}
X=\mu^{-1}(0)/\!\! /_{\theta}G\,,
\end{equation}
depending on a certain choice of stability parameter $\theta\in {\mathbb{Z}}^I$ \cite{Ginzburg:}. Group
\begin{equation}
G=\prod GL(Q_{ij})\times\prod GL({W}_i)\times \mathbb{C}^{\times}_\hbar
\end{equation}
acts as automorphisms of $X$, coming form its action on the space of representations $\text{Rep}(\mathbf{v},\mathbf{w})$ of the quiver. Here $Q_{ij}$ is the incidence matrix of the quiver, $\mathbb{C}_{\hbar}$ scales cotangent directions with character $\hbar$ and their symplectic form with $\hbar^{-1}$. We denote by $T=\mathbb{T}(G)$ the maximal torus of $G$.

For given $X$ one can define a set of tautological bundles on it $V_i, W_i, i\in I$ as bundles constructed by assigning to each point the corresponding vector spaces $V$ and $W$. Bundles $W_i$ are topologically trivial. Tensorial polynomials of these bundles and their duals generate K-theory ring $K_{\mathsf{T}}(X)$ according to Kirwan surjectivity conjecture, which is recently shown to be true on the level of cohomology \cite{McGerty:2016kir}.

\subsection{Quasimaps to Nakajima Quiver Varieties}
First let us recall the basics of the quasimap theory to quiver varieties (see \cite{Pushkar:2020aa}, Sec. 2.2 or Definition 7.2.1 of \cite{Ciocan-Fontanine:2011tg}). Let us fix a rational curve $\mathcal{D}\simeq\mathbb{P}^1$ and a set of distinct points $p_1,\dots, p_m\in\mathcal{D}$.

\begin{definition}
A  stable genus zero quasimap from curve $\mathcal{C}$ to $X$ relative to points $p_1,\dots, p_m$ is given by the following data
$$
(f,\mathcal{C},P,\pi,p'_1,\dots, p'_m)
$$
where
\begin{itemize}
\item $\mathcal{C}$ is a genus zero connected curve with at worst nodal singularities and $p'_1,\dots, p'_m$ are nonsingular points of $\mathcal{C}$.
\item P is a principal $G$-bundle over $\mathcal{C}$
\item $f$ is a section of 
$
P\times_G T^* \text{Rep}(\boldsymbol{v},\boldsymbol{w})
$
satisfying $\mu =0$, where
\item $\pi: \mathcal{C}\to\mathcal{D}$ is a regular map satisfying the following conditions:
\end{itemize}
\begin{enumerate}
\item There is a distinguished component $\mathcal{C}_0$ of $\mathcal{C}$ so that $\pi$ restricts to an isomorphism: $\pi: \mathcal{C}_0\simeq \mathcal{D}$ and $\pi(\mathcal{C}\backslash \mathcal{C}_0)$ is zero-dimensional
\item $\pi(p'_i)=p_i$
\item $f(p)$ is stable for all but a finite set of points disjoint from $p'_1,\dots, p'_m$ and the nodes of $\mathcal{C}$.
\item The line bundle $\omega_{\widetilde{\mathcal{C}}}(\sum_i p_i+\sum_k q_k)\otimes\mathcal{L}^\epsilon_\theta$ is ample for every rational positive $\epsilon$ where $\mathcal{L}^\epsilon_\theta=P\times_G \mathbb{C}_\theta$, $\widetilde{\mathcal{C}}$ is the closure of $\mathcal{C}\backslash \mathcal{C}_0$, $q_k$ are the nodal points of $\widetilde{\mathcal{C}}$ and $\mathbb{C}_\theta$ is the one dimensional $G$-module defined by the stability condition.
\end{enumerate}
\end{definition}

Conditions (1)-(3) above imply that curve $\mathcal{C}$ is a union of the distinguished component $\mathcal{C}_0$ and chains of rational curves attached to it at pints $p_1,\dots,p_m$ via a nodal singularity. Projection $\pi$ collapses the chain of $\mathbb{P}^1$s attached to each point $p_i$ to that point itself. Away from those points $\pi$ is an isomorphism. Condition (4) above states that the total number of special points on each component of $\mathcal{C}$ including the nodes is at least two. This means that point $p'i$ is located on the last component of its chain.

\begin{definition}
A relative quasimap $
(f,\mathcal{C},P,\pi,p'_1,\dots, p'_m)
$ is nonsingular at $p\in\mathcal{C}$ if $f(p)$ is stable. In this case $f(p)$ is a point in the quiver variety.
\end{definition}

\begin{definition}
The degree of a quasimap $
(f,\mathcal{C},P,\pi,p'_1,\dots, p'_m)
$ is $\boldsymbol{d}=(d_i)_{i\in\mathbb{Z}}$ where $d_i$ are degrees of rank $\boldsymbol{v}_i$ vector bundles $P\times_G \mathscr{V}_i\to\mathcal{C}$.
\end{definition}

\begin{theorem}[\cite{Ciocan-Fontanine:2011tg}]
The stack $\textsf{QM}_{{\rm relative}\, p_1,\dots,p_m}^{\ard}$ parameterizing the data of stable genus zero quasimaps to $X$ is a Deligne-Mumford stack of finite type with a perfect obstruction theory.
\end{theorem}

\begin{definition}
Let $\textsf{QM}_{{\rm nonsing}\, p_1,\dots,p_m}^{\ard}$ be the stack parameterizing the data of degree $\boldsymbol{d}$ quasimap to $X$ relative to $p_1,\dots, p_m$ such that $\mathcal{C}\simeq\mathcal{D}$. 
\end{definition}

Restricting the obstruction theory of $\textsf{QM}_{{\rm relative}\, p_1,\dots,p_m}^{\ard}$ gives a perfect obstruction theory on $\textsf{QM}_{{\rm nonsing}\, p_1,\dots,p_m}^{\ard}$.

We can define an evaluation map to the quotient stack
$$
{\rm ev}_p (f,\mathcal{C},P,\pi,p'_1,\dots, p'_m) = f(p)\in[\mu^{-1}(0)/G]\,.
$$
Given a Shur functor $\tau$ in tautological bundles on $X$ we can consider the associated K-theory class on $[\mu^{-1}(0)/G]$. This allows us to define an induced K-theory class $\tau_{{\rm stack}}$ on $\textsf{QM}_{{\rm relative}\, p_1,\dots,p_m}^{\ard}$ as
$
\tau|_p = {\rm ev}_p ^*(\tau_{{\rm stack}})\,.
$

\subsection{The Quantum K-theory}
The quasimap moduli spaces have a natural action of maximal torus, lifting its action from $X$. When there are at most two special points $0$ and  and the base curve $\mathcal{C}$ is ${\mathbb{P}}^1$ we extend $T$ by additional torus $\mathbb{C}^{\times}_q$, which scales ${\mathbb{P}}^1$ such that the tangent space $T_{0} {\mathbb{P}}^1$ has character $q$. We shall include this action in the full torus by $\mathsf{T}=T\times \mathbb{C}^{\times}_q$. We assume that the two fixed points of $\mathbb{C}^{\times}_q$ are $p_1=0$ and $p_2=\infty$. One can construct various enumerative invariants of $X$ using virtual structure sheaves $\mathscr{O}_{\text{vir}}$ for $\textsf{QM}^{\ard}$ \cite{Okounkov:2015aa}. Using the above two marked points one can define a vertex function.
\begin{definition}
The element
\begin{equation}
V^{(\tau)}(z)=\sum\limits_{\ard=\textbf{0}}^{\infty} z^{\ard} {{\rm ev}}_{p_2, *}\Big(\textsf{QM}^{\ard}_{{{\rm nonsing}} \, p_2},\widehat{{\mathscr{O}}}_{{{\rm vir}}} \tau (\left.\mathscr{V}_i\right|_{p_1}) \Big) \in  K_{\mathsf{T}}(X)_{loc}[[z]]
\label{eq:vertexQKgen}
\end{equation}
is called bare vertex with descendent $\tau\in K_T(X)$.
\end{definition}

\subsection{A-type Quivers with Framing}
For the rest of this section, we shall focus on A-type quivers, in particular, on cotangent bundles to partial flag varieties $T^*\mathbb{F}l_n$ (see Fig. \ref{fig:QuiverVarieryNak}).
Interestingly, for these spaces, our construction leads to integrable systems of two different types -- quantum spin chains (like XXZ) and dynamical many-body systems (like trigonometric Ruijsenaars-Schneider model). It has been known for some time (partly thanks to \cite{Gaiotto_2013}) that these two types of models are related by a nontrivial \textit{quantum/classical} duality. 

Our results explain this duality geometrically. It is known that K-theory vertex functions satisfy difference equations, i.e., quantum Knizhnik-Zamolodchikov equations. In \cites{Koroteev:2018azn,Koroteev:2021aa} We showed that normalized quantum K-theoretic vertex functions for cotangent bundles to partial flag varieties are the eigenfunctions of trigonometric Ruijsenaars-Schneider (tRS) Hamiltonians. Remarkably, the tRS eigenvalue problem can be formulated in two different frames -- in the first frame, the difference operators act on quantum parameters $z_i$, whereas in the second frame they act on equivariant parameters of the framing. These two frames are related to each other by \textit{3d mirror symmetry}.
In other words, tRS wave functions coincide with K-theory vertex functions and the information about the enumerative invariants can be obtained directly from integrability. The semiclassical limit $q\to 1$ will lead to equations of motion of the tRS model which will serve as relations in the ideal of quantum K-theory ring for $X$. To summarize, we proved two theorems about quantum K-theory of type-A quiver varieties:

\begin{theorem}
For a given quiver variety $X$ of type $A_r$ the eigenvalues of the quantum multiplication operator by quantum class $\widehat{\tau}(z)$ (see \eqref{eq:vertexQKgen}) are given by the character $\tau(s_{I})$ of the corresponding virtual representation of the global symmetry $GL(V)$ evaluated on the solutions of Bethe ansatz equations for the $\mathfrak{gl}_r$ XXZ spin chain. 
\end{theorem}
This theorem relies on certain non-degeneracy conditions of the Bethe roots. Upon degeneration of the form $s_i/s_j = \hbar^{\mathbb{Z}}$, many statements of \cite{Koroteev:2021tl} need to be revised. In particular, the operators of the corresponding Bethe algebra will not be diagonalizable any longer and the Bethe equations will be modified. This requires further study.

Some progress in this direction was reported in \cite{Koroteev:2020mxs}, where the degenerations of Bethe roots were used to study the embedding of any type-A quiver variety into a type-A quiver variety with all framings at the rightmost vertex of the quiver. See also \cite{Dinkins:2023aa}.

The operator of quantum multiplication by the generating function of tautological bundles in quantum K-theory of $X$ coincides with Baxter Q-operator for the XXZ chain. One identifies each energy sector of the Hilbert space of the spin chain with the space of equivariant localized K-theory of $X$. The Hilbert space of the spin chain can be regarded as a module of a quantum affine algebra for $\mathfrak{gl}_r$.

\begin{theorem}
Quantum equivariant K-theory of the cotangent bundle to complete n-flag reads
\begin{equation}
QK_T(T^*\mathbb{F}l_n)=\frac{\mathbb{C}[\zeta_1^{\pm 1},\dots,\zeta_n^{\pm 1}; a_1^{\pm 1},\dots,a_n^{\pm 1},\hbar^{\pm 1}; p_1^{\pm 1},\dots,p_n^{\pm 1}]}{(H_r(\zeta_i, p_i,\hbar)- e_r(a_1,\dots, a_n))}\,,
\label{eq:Kthflag}
\end{equation}
where $H_r$ are tRS Hamiltonians.
\end{theorem}

The $n$-particle trigonometric Ruijsenaars-Schneider model (or relativistic Calogero-Moser-Sutherland model) is an algebraic integrable system whose phase space is the moduli space of flat $GL(n;\mathbb{C})$ flat connections on a torus with one puncture. The eigenvalues of the A-cycle and B-cycle monodromy matrices describe coordinates and momenta respectively, each taking values in $\mathbb{C}^\times$. 
Both tRS coordinates and momenta have direct geometric meanings in our construction. The former are related to quantum deformation parameters $z_1,\dots, z_{n-1}$, while their canonically conjugate momenta $p_i$ are nothing but multiplication operators by quantum tautological classes $\hbar^{j-\frac{1}{2}}\widehat{\Lambda^jV_j}(z)\circledast\widehat{\Lambda^{j-1}{V^*}_{j-1}}(z)$.
It is assumed that the monodromy around the puncture has a sole distinct eigenvalue, which is equal to the equivariant parameter $\hbar$ in our construction.

\subsection{A-Type Quiver Varieties and ADHM Moduli Spaces}
In \cite{Koroteev:2021aa} we studied the quantum geometry of Nakajima quiver varieties of two different types -- framed A-type quivers and ADHM-type quivers (a typical example -- Hilbert scheme of points on $\mathbb{C}^2$). 
The follwoing equivalence between the following theories was proven:
\begin{itemize}
\item $\mathsf{T}$-equivariant K-theory of the moduli space of genus zero quasimaps to $A_n$-type quiver $X_n$ shown on the left in Fig.\ref{fig:QuiverVarieryNak}. 
\begin{equation}
\mathcal{H}_n=K_{\mathsf{T}}(\textbf{QM}(\mathbb{P}^1,X_n)). \notag
\end{equation}
We will work with $\textbf{w}_{n-1}=n$ and $\textbf{v}_{i}=i$ for $i=1,\dots,n-1$; in other words, $X_n$ is the cotangent bundle to \textit{complete} flag variety in $\mathbb{C}^n$.
In the above $\mathsf{T}$ is the maximal torus of $GL(n,\mathbb{C})\times \mathbb{C}_q^\times\times \mathbb{C}_\hbar^\times$, where $GL(\textbf{w}_i,\mathbb{C})\times \mathbb{C}^{\times}_\hbar$
acts as automorphisms of $X_n$ and $\mathbb{C}^\times_{\hbar}$ scales the cotangent directions with weight $\hbar$, while $\mathbb{C}^\times_{q}$ acts multiplicatively on the base curve.

\item $\mathbb{C}_q^\times\times \mathbb{C}_\hbar^\times$-equivariant K-theory of the ADHM moduli space 
\begin{equation}
\bigoplus\limits_{l=0}^k K_{q,\hbar}(\mathcal{M}_{\text{ADHM}})\,. \notag
\end{equation}
Geometrically $\mathcal{M}_{\text{ADHM}}$ shown on the right of Fig. \ref{fig:QuiverVarieryNak} describes the moduli space of torsion-free sheaves of rank $N$ on $\mathbb{C}^2$, where $\mathscr{V}$ is a tautological bundle of rank $l$ and $\mathscr{W}=\mathbb{C}^N$ is a trivial bundle of rank $N$. We assume that $l<n$ for all $l$. For $N=1$ this moduli space coincides with the Hilbert scheme of $l$ points on $\mathbb{C}^2$. The maximal torus $\mathbb{C}_q^\times\times \mathbb{C}_\hbar^\times$ acts naturally on $\mathbb{C}^2$ by dilations.
\end{itemize}
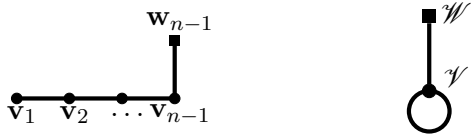
\begin{figure}[!h]
\begin{tikzpicture}[scale =0.7]
\draw [ultra thick] (0,0) -- (3,0);
\draw [ultra thick] (3,1) -- (3,0);
\draw [fill] (0,0) circle [radius=0.1];
\draw [fill] (1,0) circle [radius=0.1];
\draw [fill] (2,0) circle [radius=0.1];
\draw [fill] (3,0) circle [radius=0.1];
\node (1) at (0.1,-0.3) {$\mathbf{v}_1$};
\node (2) at (1.1,-0.3) {$\mathbf{v}_2$};
\node (3) at (2.1,-0.3) {$\ldots$};
\node (4) at (3.1,-0.3) {$\mathbf{v}_{n-1}$};
\fill [ultra thick] (3-0.1,1) rectangle (3.1,1.2);
\node (5) at (3.1,1.45) {$\mathbf{w}_{n-1}$};
\end{tikzpicture}
\qquad \qquad \qquad
\begin{tikzpicture}[xscale=0.9, yscale=0.9]
\fill [ultra thick] (1-0.1,1) rectangle (1.1,1.2);
\draw [fill] (1,0) circle [radius=0.1];
\draw [-, ultra thick] (1,1) -- (1,0.1);
\draw[-, ultra thick]
(1,0) arc [start angle=90,end angle=450,radius=.3];
\node (1) at (1.4,1.15) {$\mathscr{W}$};
\node (2) at (1.4,0.2) {$\mathscr{V}$};
\end{tikzpicture}
\caption{Left: $A_{n-1}$ quiver variety with framing on the last node (cotangent bundle to a flag variety). Right: The ADHM quiver.}
\label{fig:QuiverVarieryNak}
\end{figure} 

\begin{Thm}
\label{Th:EmbeddingTh}
For $n>k$ there is the following embedding of Hilbert spaces
\begin{align}
\bigoplus\limits_{l=0}^k K_{q,\hbar}(\text{Hilb}^l(\mathbb{C}^2)) &\hookrightarrow \mathcal{H}_n
\label{eq:Embeddingn}\\
[\lambda]&\mapsto \mathsf{V}_{\textbf{q}}\,.\notag
\end{align}
for K-theory vertex function for some fixed point $\textbf{q}$ of maximal torus $T$, where $[\lambda]$ are fixed points of $\mathbb{C}_q^\times\times \mathbb{C}_\hbar^\times$.
The statement also holds in the limit $n\to\infty$
\begin{equation}
\bigoplus\limits_{l=0}^\infty K_{q,\hbar}(\text{Hilb}^l(\mathbb{C}^2)) \hookrightarrow \mathcal{H}_\infty\,,
\label{eq:Embeddinginf}
\end{equation}
where $\mathcal{H}_\infty$ is defined as a stable limit of $\mathcal{H}_n$ as $n\to\infty$.
\end{Thm}

\vspace{2mm}

The study of quantum geometry can be developed further along these lines; some of the current research directions are outlined below.

\subsection{DT/GW Correspondence}
In \cite{Koroteev:2021aa} a conjecture which relates spectra of quantum multiplication operators in K-theory of the ADHM moduli spaces with the solution of the \textit{elliptic} Ruijsenaars-Schneider model was made. 
This conjecture is partly motivated by the correspondence between quantum geometry of the ADHM moduli spaces and Donaldson-Thomas theory on $\mathbb{P}^1\times\mathbb{C}^2$, where all the results so far are obtained in cohomology. We showed that spectra of the tRS Hamiltonians are in one-to-one correspondence with \textit{classical} multiplication by universal bundles in the equivariant K-theory of the ADHM moduli space. Therefore, we can make a conjecture that eigenvalues of \textit{quantum} multiplication will be given by \textit{elliptic} RS Hamiltonians:

\begin{Con}\label{Conj:eRSEigenproblem}
The eigenfunction of elliptic Ruijsenaars-Schneider Hamiltonians
\begin{equation}
H_r(\textbf{x})=\sum_{\substack{\mathcal{I}\subset\{1,\dots,n\} \\ |\mathcal{I}|=r}}\prod_{\substack{i\in\mathcal{I} \\ j\notin\mathcal{I}}}\frac{\theta_1(\hbar x_i/x_j|\mathfrak{p})}{\theta_1(\hbar x_i/x_j|\mathfrak{p})}\prod\limits_{i\in\mathcal{I}}p_k \,,
\label{eq:eRSRelationsEl}
\end{equation}
where $\mathfrak{p}\in\mathbb{C}^\times$ is the new parameter which characterizes the elliptic deformation away from the trigonometric locus,
is given by the K-theoretic holomorphic equivariant Euler characteristic of the affine Laumon space 
\begin{equation}
\mathcal{Z} = \sum_{\textbf{d}} \mathfrak{q}^{\textbf{d}} \int\limits_{\mathcal{L}_{\textbf{d}}} 1\,,
\label{eq:equivaraintKthLaumon}
\end{equation}
where $\mathfrak{q}=(\mathfrak{q}_1,\dots,\mathfrak{q}_n)$ is a string of $\mathbb{C}^\times$-valued coordinates on the maximal torus of $\mathcal{L}^{\text{aff}}_{\textbf{d}}$.
The eigenvalues $\mathscr{E}_r$ are equivariant Chern characters of bundles $\Lambda^r \mathscr{W}$, where $\mathscr{W}$ is the constant bundle of the corresponding ADHM space.
\end{Con}

For instance, the operator of multiplication by universal bundle
\begin{equation}
\mathscr{U} = \mathscr{W}+(1-\hbar^{-1})(1-q)\mathscr{V}
\end{equation}
in quantum equivariant K-theory $K_{q,\hbar}(\text{Hilb}^k(\mathbb{C}^2))$ is given by
\begin{equation}
\mathcal{E}_1(\lambda) = 1-(1-\hbar^{-1})(1-q) (s_1+\dots+s_k)\,,
\label{eq:EUnivBundleEigen}
\end{equation}
 $s_1,\dots s_k$ which solve Bethe ansatz equations for the ADHM quiver.

The conjecture was verified using \textrm{Mathematica} up to several orders in expansion in the elliptic parameter $p$. 
Section 5 of \cite{Koroteev:2021aa} contains a plethora of evidence for this conjecture.
Some further development in this direction was made by the author and Zeitlin in \cite{Koroteev:2020mxs} in the context of toroidal $q$-opers, which will be described in the next section.

\section{Opers}\label{Sec:Opers}

\subsection{Baxter $Q$-Operators}
It is known that the {\it universal R-matrix}, which produces particular braiding operators $R_{V_i(a_i), V_j(a_j)}$ belongs to the completion of the tensor product $U_{\hbar}(\hat{\mathfrak{b}}_{+})\otimes U_{\hbar}(\hat{\mathfrak{b}}_{-})$, where $U_{\hbar}(\hat{\mathfrak{b}}_{\pm})$ are the Borel subalgebras of $U_{\hbar}(\hat{\mathfrak{g}})$. Thus, one can take auxiliary representations $W(u)$ in the definition of the transfer matrix \eqref{eq:TransfermatDef} to be representations of $U_{\hbar}(\hat{\mathfrak{b}}_{+})$. 

There exist prefundamental representations of $U_{\hbar}(\hat{\mathfrak{b}}_{+})$, which are infinite-dimensional. If one extends the braided tensor category of finite-dimensional modules by such representations, it appears that prefundamental representations generate the entire category. 

The corresponding transfer matrices turn out to be well-defined and 
moreover, the eigenvalues of the transfer matrices are polynomials of the spectral parameter, generating elementary symmetric functions of the solutions of Bethe equations. 
Such transfer matrices are known as {\it Baxter operators}.

There are two series of prefundamental representations $\{V^i_+(u)\}_{i=1,\dots, r}$, $\{V^i_-(u)\}_{i=1,\dots, r}$ and the associated Baxter operators $\{Q^i_{\pm}(u)\}_{i=1,\dots, r}$.  They obey the following key relations
\begin{equation}
\label{eq:QQSystem}
{{\widetilde\xi_i}}Q^i_{-}({u})Q^i_{+}({\hbar}  {u})-{\xi_i}Q^i_{-}({\hbar}  {u})Q^i_{+}({u}) =\Lambda_i({u})\prod_{j\neq i}\Bigg[\prod^{-a_{ij}}_{k=1} Q^j_{+}({\hbar} ^{b_{ij}^k}{u})\Bigg]\,,\quad i=1,\dots,r, \quad b_{ij}^k\in \mathbb{Z}
\end{equation}
Here, polynomials $\Lambda_i({u})$ are known as Drinfeld polynomials, characterizing the representation $\mathcal{H}$ of $U_{{\hbar}}(\hat{\mathfrak{g}})$ and 
${\xi_i}$, ${{\widetilde\xi_i}}$ are some monomials of $\{z_i\}$.
This system of equations, known as the {\it $QQ$-system}, considered as equations on  $\{Q^i_{\pm}(u)\}_{i=1,\dots, r}$ and subject to some non-degeneracy conditions, are equivalent to the Bethe ansatz equations.

\subsection{Miura $\hbar$-opers}
Let $^L G$ be the simply-connected group with Lie algebra $^L \mathfrak{g}$. There is a natural classical object, the $\hbar$-difference connection, locally a meromorphic $^L G$-valued function $A(z)$ on Zariski open set of $\mathbb{P}^1$, which transform upon trivialization change $A(z)\to g(\hbar z)A(z)g^{-1}(z)$.

Together with Frenkel, Sage and Zeitlin \cite{Frenkel:2020}, following the constructions in \cite{KSZ} done for $SL(N)$,  
we developed the $\hbar$-difference analog of opers as such $\hbar$-difference connections for any simply-connected semisimple Lie group $^L G$ with a fixed Borel subgroup $^L B_-$ (see more recent review \cite{Zeitlin:2024zvs}).  
Locally, these $\hbar$-connections have the form $A(z)=n'(z)\prod^r_{i=1}s_i\phi_i^{\check{\alpha}_i}(z)n(z)$. Here $n(z), n'(z)\in G(z)$, $\phi_i(z)\in \mathbb{C}(z)$, $s_i$ are the lifts of the fundamental Weyl reflections to $^L G$. In other words,  $A(z)\in B_-(z)cB_-(z)$, where $c=\prod^r_{i=1}s_i$ is a Coxeter element.

Moreover, we defined such $(^L G, \hbar)$-opers and their Miura versions with regular singularities, which amounts to the connections of this type that preserve the opposite Borel subgroup of $B_+$ and taking $\phi_i(z)=\Lambda_i(z)\in \mathbb{C}[z]$.  We proved several structural theorems about them.

One of the major statements we make in \cite{Frenkel:2020} is devoted to the explicit relation of these objects to the $QQ$-systems and Bethe ansatz. To do that, we work with two versions of what we call {\it Z-twisted condition} for Miura opers. The simplest $Z$-twisted condition implies that the $(^L G,\hbar)$-oper connection can be $\hbar$-gauge equivalent to the semisimple element $Z\in H\subset ^L G$,  where $H$ is the Cartan subgroup. That means  $A(z)=g(\hbar z)Zg^{-1}(z)$. This condition is a different version of the zero monodromy condition and double pole irregular singularity at $\infty$ point of $\mathbb{P}^1$.

The relaxed version of this $Z$-twisted condition is as follows. Given the principal $^L G$-bundle, one can construct an associated bundle for any fundamental representation  $V_{\omega_i}$ for the fundamental weight $\omega_i$.  
One can associate a $(GL(2),\hbar)$-oper to any such pair  $(^L G,\hbar)$-oper and $V_{\omega_i}$: this is done by restricting the Miura $(^L G,\hbar)$-oper to the two-dimensional subspace, spanned by two top weights in $V_{\omega_i}$. This is possible, since Miura $(^L G,\hbar)$-oper preserves the reduction to positive Borel subgroup $^L B_+\subset ^L\!\! G$.  

We say that the resulting Miura oper is {\it $Z$-twisted Miura-Pl\"ucker}  $(^L G,\hbar)$-oper if for every such $(GL(2),\hbar)$ oper is $\hbar$-gauge equivalent to the restriction of $Z$ to the corresponding two-dimensional space. 

In \cite{Frenkel:2020} we showed that {\it $Z$-twisted Miura-Pl\"ucker}  $(^L G,\hbar)$-opers with mild non-de\-ge\-ne\-racy conditions are in one-to-one correspondence with certain $QQ$-systems and that does not depend on the order in the Coxeter element. In simply-laced cases such $QQ$-systems are equivalent to standard Bethe ansatz equations. The non-simply laced case is more involved and needs to be properly understood.

Our results suggest that when $G=SL(N)$ then the duality between the space of $(^L G,\hbar)$-opers and the space of solutions of $QQ$-systems
\eqref{eq:QQSystem} (or generalization of the quantum/classical duality of \cite{Gaiotto_2013}, where $\hbar$-opers were represented by the phase space of the tRS model) is equivalent to the classical (critical level) version of the quantum geometric Langlands correspondence. Yet, beyond Type A there is a reason to believe that such a connection still exists, albeit it more subtle.

We suspect that these equations for the non-simply laced cases can be obtained from twisted root systems of $^L\mathfrak{g}$.

\vskip.1in

While it immediately follows that any $Z$-twisted Miura oper is indeed  $Z$-twisted Miura-Pl\"ucker oper, the opposite statement, however, is highly nontrivial. In \cite{Frenkel:2020} we introduce a chain of $\hbar$-gauge  transformations, which we refer to as $\hbar$-B\"acklund transformations, which on the level of $QQ$-systems amounts to the $Q^i_+(z)\mapsto Q^i_-(z)$, $Z\mapsto s_i(Z)$, where $s_i$ is elementary Weyl reflection. However, at every step, in order to progress further, we have to impose the non-degeneracy condition on the $QQ$-system and the associated Miura oper. We have shown that if one can proceed with this transformation to $Z$-twisted Miura-Pl\"ucker oper, corresponding to the $w_0(Z)$, where $w_0$ is the longest Weyl group element, then such $Z$-twisted Miura-Pl\"ucker Miura oper is $Z$-twisted. We call such Miura-Pl\"ucker opers and the associated $QQ$-system $w_0$-generic. 


\section{Some Open Problems}\label{Sec:OpenProblems}
In this section, we shall address some open problems which have recently appeared at the interface between enumerative geometry and integrability.

\subsection{Compact Limit}\label{Sec:CompatLimit}
As discussed in \secref{Sec:QuantumIntEnum}, the equivariant K-theory of Nakajima quiver varieties contains parameter $\hbar$ which acts on the fiber by dilations. In the $\hbar\to\infty$ limit the equivariant volume of the fiber vanishes and the cotangent bundle retracts to its zero section. 
Theorem \ref{Th:EmbeddingTh}, when both sides of the duality undergo this limit, deserves separate attention. In \cite{Koroteev:2018azn}, A. Zeitlin and the author demonstrated that Givental-Lee's J-functions of quantum K-theory of flag varieties are directly reproduced from $\mathsf{V}_{\textbf{q}}$. On the Hilbert scheme side, the eigenvalue \eqref{eq:EUnivBundleEigen} of quantum multiplication becomes
\begin{equation}
\mathscr{E}_1(\lambda) = 1-(1-q) (s_1+\dots +s_k)\,,
\label{eq:EUnivBundleEigennew}
\end{equation}
where Bethe roots $s_i$ now solve \begin{equation}
\prod_{l=1}^N(s_a-\mathrm{a}_l)\cdot\prod_{\substack{b=1 \\ b\neq a}}^k\frac{s_a-q s_b}{s_a-q^{-1}s_b}=\mathfrak{p}^{\Lambda}\,,\quad a=1,\dots, k\,,
\label{eq:BetheADHMnew}
\end{equation}
where $\mathfrak{p}^{\Lambda}$ is a new parameter. The $\hbar\to\infty$ limit effectively reduces the moduli space of sheaves to the Hilbert scheme of $k$ points on single $\mathbb{C}$ (which is merely a symmetric product of $k$ copies of $\mathbb{C}$ in this case).
Yet our results (as well as string theory constructions) suggest that this space -- the moduli space $\mathcal{M}_k^{\text{vort}}$ of $k$ vortices. In physics literature, the above equations describe \textit{vortex} moduli spaces (as opposed to instanton moduli spaces in case of ADHM).  On $\mathbb{C}$ --  should have a nontrivial structure of quantum multiplication.
One should be able to reproduce the above result  by a direct enumerative computation for the quantum multiplication \eqref{eq:EUnivBundleEigennew}.

Recently, an alternative description of quantum K-theory of partial flag varieties has been proposed in \cites{Gu:2022aa,Gu:2023aa}, which does not utilize Toda system coordinates. Their description contains more relations (so-called Whitney relations) and more generators. In particular, unlike the Toda case, where we only uses products of top degree exterior powers of the tautological bundles $\Lambda^i V_i$ and their duals, Gu et. al.'s work also involves lower wedge powers and hence more relations. It would be interesting to understand the connection between the two approaches.

It would be interesting to find the connection between our approach to equivariant K-theory of K\"ahler spaces (partial flags, etc.) and recent developments by Gu et. al.
\label{eq:MihalceaetalKth}

\subsection{Open Spin Chains and Isotropic Quiver Varieties}\label{Sec:OpenChainsSpSO}
There exists a new class of quiver varieties, called $\sigma$-quiver varieties, turns out to be fixed-point subvarieties of Nakajima varieties under certain symplectic involutions \cite{Li:2018aa}.

Opers of orthogonal and symplectic types are interesting from various perspectives. From the integrable systems vantage point, one expects to find a number of quantum/classical dualities between spin chains of $B_n$ (or $C_n$) type with Calogero-Ruijsenaars systems of Langlands dual $C_n$ (or $B_n$) types respectively. Additionally, one is expected to discover spin chains with open boundary conditions.

As in the (q)-oper construction, consider automorphism $M_q: \mathbb{P}^1\to\mathbb{P}^1$ acting as $z\mapsto q z$ for $q\in\mathbb{C}^\times$ together with $\mathbb{Z}_2$ reflection $T$ acting as $z\mapsto z^{-1}$. The idea is to study $T$-invariant supbspaces of the spaces of $GL(n)$ opers.

We expect that the spaces of such `orbifolded' opers will be related to quantum cohomology and quantum K-theory of \textit{isotropic} quiver varieties, see \cite{Li:2018aa} for the recent work.

\vskip.1in
The oper construction which we reviewed in \secref{Sec:Opers} can be applied to the open chain set-up as well. Consider automorphism $M_q: \mathbb{P}^1\to\mathbb{P}^1$ acting as $z\mapsto q z$ for $q\in\mathbb{C}^\times$ together with $\mathbb{Z}_2$ reflection $T$ acting as $z\mapsto z^{-1}$. 

We shall first consider an example of an orbifolded $(GL(2),q)$-oper. For a Zariski open dense subset $U\subset\mathbb{P}^1$ consider $V=U\cap M_q^{-1}(U)$. A meromorphic orbifolded $(GL(2),q)$-oper on $\mathbb{P}^1$ is a triple $(E,A,\mathcal{L})$, where $E$ is a rank-two holomorphic vector bundle on $\mathbb{P}^1$ such that it is isomorphic to its pullback $E\simeq E^{q,T}$ with respect to $M_q\circ T$, $\mathcal{L}\subset E$ is a line subbundle, and
$A\in\text{Hom}_{\mathcal{O}_{\mathbb{P}^1}}(E,E^q)$, where $E^q$ is a pullback of $E$ under $M_q$ such that the restriction 
\begin{equation}
\label{eq:orbopdef}
\bar{A}: \mathcal{L}\to \left(E/\mathcal{L}\right)^{q}
\end{equation}
is an isomorphism on $V$. 

As being explored in an upcoming work, this construction will yield modifield $Q$-functions and Bethe ansatz equations. In particular, meromorphic sections of the oper bundle will now have the folliwing form which respects the $T$ operation
$$
Q_+(z)=\prod_{i=1}^k(z-s_i)\left(\frac{q}{z}-s_i\right)\,,
$$ 
while the corresponding Bethe equations will reproduce those of Sklyanin for an open spin chain with the special choice of the boundary $K$-operator \cite{EKSklyanin_1988}. At the moment, the main challenge is to find proper geometric description of various $K$-matrices which arize from repersentation theory.


\begin{figure}
\includegraphics[scale=0.26]{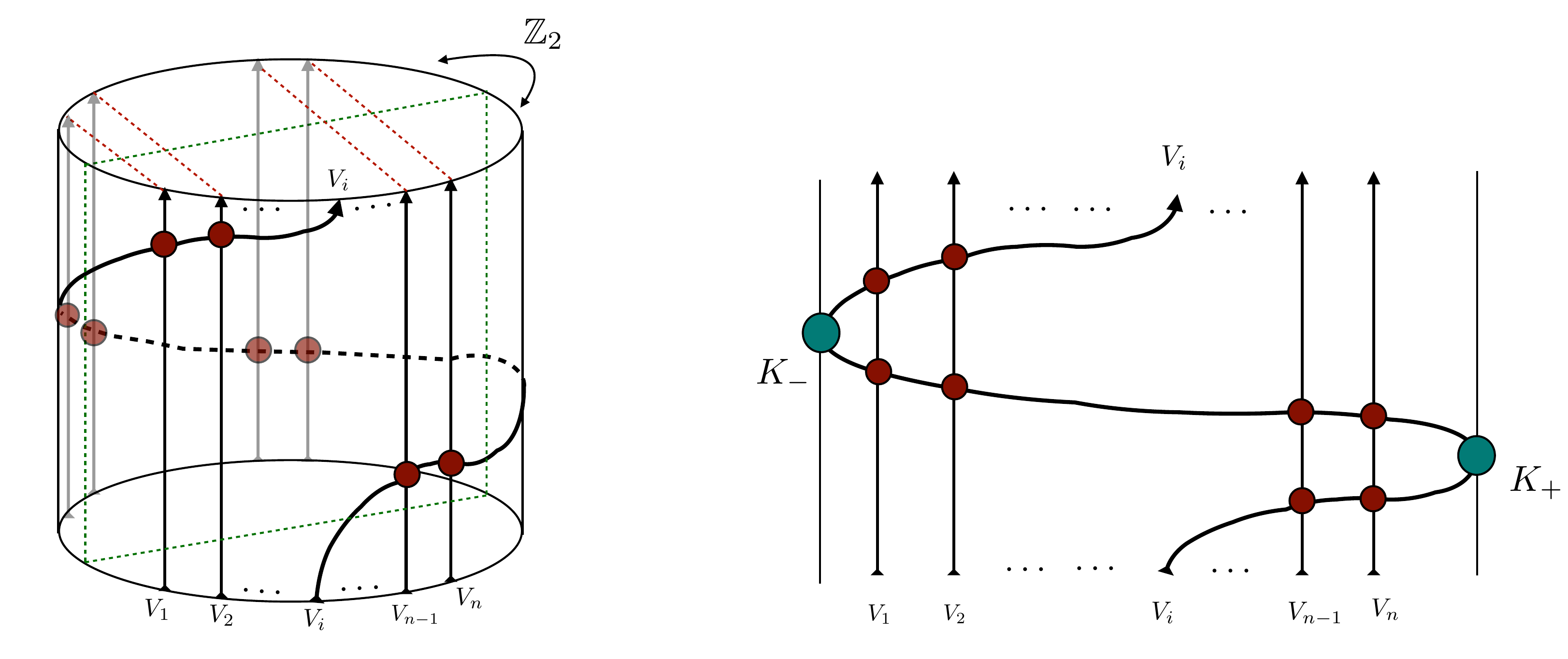}
\caption{Orbifolding trick yields open spin chain from the closed one.}
\end{figure}

\subsection{Number Theory Applications}\label{Sec:NumberTheory}
Consider equivariant cohomology of quiver variety $X$. Its vertex reads
\begin{equation}
    V(z)=\sum_{\textbf{d}>0} c_{\textbf{d}} z^{\textbf{d}}\in \mathbb{Q}[[z]]\,,
\end{equation}
where $c_{\textbf{d}}$ are regularized integrals over the virtual fundamental class of the compactified moduli space of quasimaps intro $X$.
The works of A. Smirnov and A. Varchenko \cites{Smirnov:2023aa,Smirnov:2023ab} initiated the study of arithmetic properties of these coefficients. Using 3d mirror quiver variety $X'$, it is possible to construct a system of polynomials $T_s(z)$ 
with integral coefficients which converges to the vertex function in the $p$-adic norm
\begin{equation}
    \lim_{s\to\infty} T_s(z) = V(z)
\end{equation}
These polynomials satisfy an interesting set of relations known in number theory as Dwork congruences, which
laid the foundation of the theory of $p$-adic hypergeometric equations.

Using methods of enumerative geometry, we expect to find new results which otherwise would have been difficult to obtain using purely number-theoretic methods. In \cite{Smirnov:2023aa} the vertex function for an $A_1$ quiver variety $X$ was shown to be equal to the following contour integral
\begin{equation}
V(\zeta) = \oint\limits_\gamma \Phi(x,\zeta)d\xi\,.
\end{equation}
which is an integral of a $z$-dependent meromorphic form over a cycle on a (some cover of) a (hyper)elliptic curve.  In order to find a differential operator that annihilates $V(x)$ we need to find a linear combination of $z$-derivatives, with coefficients in $\mathbb{C}(x,\zeta)$, which upon acting on the integrand $\Phi(x,\zeta)$ yields a total $x$-derivative. In a work in progress we demonstrate that this is a Calogero-Moser operator.


%

\bibliography{cpn12}
\end{document}